\documentclass[a4paper,11pt]{article}
\pdfoutput=1 

\usepackage{jheppub} 

\usepackage[T1]{fontenc} 
\usepackage{comment}
\usepackage{tikz}
\usepackage{amsmath,amssymb}

\title{Chiral higher-spin holography in flat space: the Flato-Fronsdal theorem and lower-point functions}

\author{Dmitry Ponomarev}

\affiliation{Institute for Theoretical and Mathematical Physics,\\
Lomonosov Moscow State University, Lomonosovsky avenue, Moscow, 119991, Russia}
\affiliation{I.E. Tamm Theory Department, Lebedev Physical Institute,\\
 Leninsky avenue, Moscow, 119991, Russia}

\emailAdd{ponomarev@lpi.ru}

\abstract{
We prove the flat space analogue of the Flato-Fronsdal theorem. It features the flat space singleton representation suggested recently. We do that by deriving a kernel that intertwines a pair of singleton representations with massless higher-spin fields in flat space. Next, we derive two-point functions of flat space singletons, which are then used to construct two- and three-point scattering amplitudes in the dual theory of massless higher-spin fields. These amplitudes agree with amplitudes in the chiral higher-spin theory. 
}

\begin{document} 
\maketitle
\flushbottom

\section{Introduction}

Flat space holography attracted considerable attention in recent years, see e. g. \cite{Strominger:2017zoo,Pasterski:2021rjz,Prema:2021sjp,McLoughlin:2022ljp} for review. Despite significant progress in this area, explicit holographically dual pairs with both theories independently defined are lacking,  see, however, recent suggestions \cite{Costello:2022jpg,Stieberger:2022zyk,Rosso:2022tsv}. As was  reviewed in \cite{Ponomarev:2022ryp}, this difficulty originates from the fact, that  flat space counterparts of boundary conformal field theories with the desired properties do not exist. In particular, ultra-relativistic contractions of conformal field theories lead to nilpotent translations, which is incompatible with the non-trivial action of translations for the usual fields in the Minkowski bulk space. This suggests that the familiar holographic setup cannot be directly carried over from the AdS space to the flat space, instead, it requires considerable modifications.

To understand how exactly flat space holography may work a natural starting point is the setting of higher-spin holography. In the simplest scenario of the higher-spin holography in AdS \cite{Sezgin:2002rt,Klebanov:2002ja} the boundary theory is free. Due to that, to large extent higher-spin holography reduces to simple representation theory manipulations and, moreover, both sides of the duality are severely constrained by rich higher-spin symmetry. Thus, higher-spin holography in the AdS space crucially relies on a set of  simple and rigid structures, which may be used to control consistency of the deformation of the whole construction into the flat space bulk. 

Higher-spin theories in flat space are  well understood in the chiral (self-dual) sector, see \cite{Metsaev:1991mt,Metsaev:1991nb,Ponomarev:2016lrm} for original results and \cite{Ponomarev:2022vjb} for a detailed review\footnote{Some recent developments on chiral higher-spin theories can be found in \cite{Ponomarev:2017nrr,Skvortsov:2018jea,Krasnov:2021nsq,Sharapov:2022faa,Sharapov:2022awp}. Chiral higher-spin theories and related structures has been also recently discussed in the context of celestial holography \cite{Ren:2022sws,Monteiro:2022lwm,Bu:2022iak,Guevara:2022qnm}. We would also like to mention a  related earlier work \cite{Devchand:1996gv}.}. In particular, the associated global higher-spin symmetries were established in \cite{Ponomarev:2017nrr,Krasnov:2021nsq,Krasnov:2021cva,Skvortsov:2022syz,Sharapov:2022faa,Ponomarev:2022atv}. Following the standard steps this symmetry  algebra allows one to identify the dual theory, referred to as  the flat space singleton. In \cite{Ponomarev:2022ryp} it was found that the flat space singleton carries a representation of a certain  deformation of the Poincare algebra --  which can be connected to the so-called Maxwell algebra
\cite{Bacry:1970ye,Bacry:1970du,Schrader:1972zd,Bonanos:2008ez,Gomis:2017cmt}
  -- and explicit realisations of this representation were given.  This deformation of the Poincare algebra is what makes the given setup very different from more standard  ones -- usually, based on the BMS symmetry \cite{Bondi:1962px,Sachs:1962wk} -- and what allows one to overcome the difficulty with the existence of the dual theory we alluded to above. 

In the present paper we further elaborate on the proposal made in \cite{Ponomarev:2022ryp}. In particular, we show that the flat space singletons feature a version of the Flato-Fronsdal theorem \cite{Flato:1978qz}. To be more precise, we show that the tensor product of two flat space singletons decomposes into the direct sum of  representations, for which the aforementioned deformation of the Poincare algebra trivialises and which can be identified with the usual massless higher-spin fields in the Minkowski space. Our argument is based on the explicit kernel, which intertwines the  representations as stated above.

Next, we construct two-point functions of singleton representations, by requiring that these are invariant with respect to the deformed Poincare algebra.
Then, assuming that the theory of singletons is free, we compute simple correlators of single-trace operators  by performing the Wick contractions. Finally, employing the intertwining kernel we constructed previously, these give rise to two- and three-point amplitudes of the higher-spin gauge theory. The latter agree with amplitudes in the chiral higher-spin theory.

This paper is organised as follows. In the next section we review how the relevant deformed Poincare algebras are defined, present their singleton representations and the representation associated with the tower of massless higher-spin fields. In section \ref{sec:3} we construct the intertwining kernel between the tensor product of two singletons and massless higher-spin fields. In this way we prove the flat space version of the Flato-Fronsdal theorem. We then compute correlators in the theory of free singletons in section \ref{sec:4}, which are then reinterpreted as higher-spin scattering amplitudes. We conclude in section \ref{sec:5}.

\section{Deformed Poincare algebra and its representations}

In this section we recall how the relevant deformations of the Poincare algebra are defined. We next review how the massless higher-spin fields transform with respect to these algebras as well as give their singleton representations.

The deformed chiral Poincare algebra consists of the anitholomorphic half of the Lorentz transformations, $\bar J_{\dot\alpha\dot\beta}$, deformed translations $P_{\alpha\dot\alpha}$ and the central element $L_{\alpha\alpha}$. Unlike for the usual Poincare algebra deformed translations do not commute 
 \begin{equation}
\label{19sep3}
[P_{\alpha\dot\alpha},P_{\beta\dot\beta}]=-i\epsilon_{\dot\alpha\dot\beta}L_{\alpha\beta}.
\end{equation}
The dotted spinor indices refer to the complex conjugate representation of $sl(2,\mathbb{C})$ generated by $\bar J_{\dot\alpha\dot\alpha}$. Unless the remaining Lorentz generators $J_{\alpha\alpha}$ are added the undotted spinor indices do not transform, they only enumerate different components of $P_{\alpha\dot\alpha}$ and $L_{\alpha\alpha}$.
If $J_{\alpha\alpha}$ are added, the undotted indices transform in the fundamental representation of $sl(2,\mathbb{C})$. Then $L$ commutes non-trivially with $J$, so $L$ is no longer a central element. The latter algebra will be referred to as the $L$-deformed Poincare algebra and it is a version of the so-called Maxwell algebra \cite{Bacry:1970ye,Bacry:1970du,Schrader:1972zd,Bonanos:2008ez,Gomis:2017cmt}. For more details on the definitions and for explicit commutation relations, see \cite{Ponomarev:2022ryp}.

\subsection{Massless higher-spin fields}

Massless positive energy higher-spin representations of the deformed chiral Poincare algebra are given by
\begin{equation}
\label{3oct1}
\bar J_{\dot\alpha\dot\alpha} = 2i\bar\lambda_{\dot \alpha}\frac{\partial}{\partial\bar\lambda^{\dot\alpha}}, \qquad
P_{\alpha\dot\alpha}=-\lambda_\alpha\bar\lambda_{\dot\alpha}, \qquad L_{\alpha\alpha}=0.
\end{equation}
The representation space is carried by functions $\Phi(\lambda;\bar\lambda)$ on $\mathbb{C}^2/\{0\}$.
Together with the remaining part of the Lorentz algebra, which is realised as
\begin{equation}
\label{3oct2x01}
J_{\alpha\alpha}=2i\lambda_{\alpha}\frac{\partial}{\partial \lambda^\alpha}
\end{equation}
we obtain the usual massless higher-spin representations of the Poincare algebra. Indeed, due to the fact that for these representations  $L=0$, deformation (\ref{19sep3}) becomes trivial and we end up dealing with the usual massless fields. Note that once we keep only generators (\ref{3oct1}), states with different $\lambda$ are not mixed by the algebra. Then $\lambda$ can be regarded as a representation label in the sense of labelling different representations.

\subsection{Singleton representations}

For our purposes it is more convenient to use the single-variable realisation of the singleton representations given in \cite{Ponomarev:2022ryp}. To be more precise, we take
\begin{equation}
\label{24aug2}
\begin{split}
\bar J_{\dot 1\dot 1}= a^2, \qquad \bar J_{\dot 2\dot 2}=-\frac{\partial^2}{\partial a^2}, \qquad
\bar J_{\dot 1\dot 2}&= -i (a\frac{\partial}{\partial a}+\frac{1}{2}),\\
P_{\alpha\dot 1}=-\frac{1}{\sqrt{2}} \lambda_\alpha a, \qquad  P_{\alpha\dot 2}&=\frac{i}{\sqrt{2}} \lambda_\alpha \frac{\partial}{\partial a},\\
L_{\alpha\alpha} &= \frac{1}{2}\lambda_\alpha \lambda_{\alpha}.
\end{split}
\end{equation}
The representation space is given by functions $c_\lambda(a)$, where $\lambda$ labels different representations. Representation (\ref{24aug2})
can be extended to the representation of the $L$-deformed Poincare algebra. Then $J$ acts as in (\ref{3oct2x01}). Moreover, the representation space splits into two invariant subspaces defined by
\begin{equation}
\label{8oct1}
c_{-\lambda}(-a)=\pm c_\lambda(a).
\end{equation}
The even subspace is analogous to the scalar singleton in the AdS space, while the odd one is analogous to the AdS spinor singleton
\begin{equation}
\label{23nov1}
\begin{split}
\text{scalar:}& \qquad c_{-\lambda}(-a)= c_\lambda(a),\\
\text{spinor:}& \qquad c_{-\lambda}(-a)= -c_\lambda(a).
\end{split}
\end{equation}

\section{Flat space Flato-Fronsdal theorem}
\label{sec:3}

We start by giving a simple qualitative argument, which suggests how the Flato-Fronsdal theorem may work in flat space. As we could see in the previous section, massless higher-spin fields decompose into the direct sum of representations of the deformed chiral Poincare algebra labelled by  $\lambda$. At fixed $\lambda$ the representation space is given by a function of two variables $\bar\lambda^{\dot 1}$ and $\bar\lambda^{\dot 2}$. In turn, the singleton representation at fixed $\lambda$ is realised by a function of a single variable $a$. This superficial analysis suggests that the tensor product of two singletons with fixed $\lambda$ may give a massless higher-spin representation with fixed $\lambda$.

Below we will show that this is, indeed, the case. This will be done by providing an explicit intertwining map 
\begin{equation}
\label{3oct2}
C_{\lambda_1(\lambda),\lambda_2(\lambda)}(a_1,a_2)=\int d\bar\lambda^{\dot 1}d\bar\lambda^{\dot 2}
k(a_1,a_2;\bar\lambda^{\dot 1},\bar\lambda^{\dot 2})
\Phi(\lambda;\bar\lambda^{\dot 1},\bar\lambda^{\dot 2}),
\end{equation}
where $C$ is the wave function of the tensor product of two singletons and $\lambda_i$ are  uniquely defined by $\lambda$, as anticipated by
 the argument given above. The fact that the intertwining kernel $k$ is invariant under the deformed chiral Poincare algebra implies that for any of its generators $O$ one has
\begin{equation}
\label{3oct3}
O_C k = O_\Phi^T k.
\end{equation}
Here the subscripts $C$ and $\Phi$ refer to the action of $O$ in the tensor product of singletons and in the higher-spin representations, while the superscript $T$ refers to integration by parts.

\subsection{Intertwining $\bar J$}

We will start by finding the most general intertwining kernel, which is invariant with respect to the antiholomorphic half of the Lorentz algebra generated by $\bar J$.
In principle, this can be done by explicitly solving differential equations resulting from (\ref{3oct3}). This turns out to be somewhat cumbersome, moreover, by proceeding this way one may easily overlook distributional solutions. Below, we will use a different method, which relies on treating representations as the lowest-weight ones. 

For representation (\ref{24aug2}) it is suggestive to regard $i\bar{J}_{\dot 1\dot 2}$ as a weight, which is lowered by $\bar J_{\dot 2\dot 2}$ and raised by $\bar J_{\dot 1\dot 1}$. Then, the singleton representation (\ref{24aug2}) is the lowest-weight representation with the lowest-weight space spanned by $c^1_\lambda(a)=1$, $c^2_\lambda(a)=a$ and the associated eigenvalues of $i\bar{J}_{\dot 1\dot 2}$ are equal to $\frac{1}{2}$ and $\frac{3}{2}$ respectively.
The tensor product of two such representations is again a lowest-weight representation. Its lowest-weight space can be found by directly solving the lowest-weight condition
\begin{equation}
\label{3oct4}
-\left(\frac{\partial^2}{\partial a_1^2}+\frac{\partial^2}{\partial a_2^2}\right)C^{\rm lw}_{\lambda_1,\lambda_2}(a_1,a_2)=0.
\end{equation}
This leads to
\begin{equation}
\label{3oct5}
C^{\rm lw}_{\lambda_1,\lambda_2}(a_1,a_2)= \sum_{k=0}\left[ \alpha_k (a_1+ia_2)^k+\beta_k(a_1-i a_2)^k\right].
\end{equation}
The associated eigenvalues of $i\bar{J}_{\dot 1\dot 2}$ are
\begin{equation}
\label{1sep7}
\begin{split}
i\bar J_{\dot 1\dot 2}(a_1+ia_2)^k=(k+1)(a_1+i a_2)^k,\\
i\bar J_{\dot 1\dot 2}(a_1-ia_2)^k=(k+1)(a_1-i a_2)^k.
\end{split}
\end{equation}
In other words, we find that the tensor product of two singleton representations is a direct sum of lowest-weight representations of $\bar J$ with lowest weights $1,2,\dots$, each representation appearing twice. Without loss of generality, we can focus on the first branch of solutions in (\ref{3oct5}), while the second one can be obtained by permuting $a_1$ and $a_2$.
Once the lowest weight states are known, the remaining states in the representation can be obtained by applying the raising operator. Explicitly, one finds
\begin{equation}
\label{3oct6}
(\bar J_{\dot 1\dot 1})^n  (a_1+ia_2)^k = (a_1^2+a_2^2)^n (a_1+ia_2)^k = (a_1+ia_2)^{n+k}(a_1-ia_2)^n.
\end{equation}

Next, we will repeat a similar analysis for the massless higher-spin representations. The $\bar J$ generators explicitly read
\begin{equation}
\label{3oct7}
\begin{split}
\bar J_{\dot 1\dot 1}&=-2i \bar\lambda^{\dot 2}\frac{\partial}{\partial \bar\lambda^{\dot 1}},
\\
\bar J_{\dot 1\dot 2}&=-i \bar\lambda^{\dot 2}\frac{\partial}{\partial \bar\lambda^{\dot 2}}+ i\bar\lambda^{\dot 1}\frac{\partial}{\partial \bar\lambda^{\dot 1}},\\
\bar J_{\dot 2\dot 2}&=2i \bar\lambda^{\dot 1}\frac{\partial}{\partial \bar\lambda^{\dot 2}}.
\end{split}
\end{equation}
Accordingly, the lowest-weight space satisfies 
\begin{equation}
\label{1sep8}
2i \bar\lambda^{\dot 1}\frac{\partial}{\partial \bar\lambda^{\dot 2}} \Phi^{\rm lw}(\lambda;\bar\lambda^{\dot 1},\bar\lambda^{\dot 2})=0.
\end{equation}
This equation admits two branches of solutions
\begin{equation}
\label{1sep9}
\Phi^{\rm lw}(\lambda;\bar\lambda^{\dot 1},\bar\lambda^{\dot 2}) = \delta(\bar\lambda^{\dot 1})\gamma_1(\bar\lambda^{\dot 2})+\gamma_2(\bar\lambda^{\dot 1}).
\end{equation}
These two branches  are related by the Fourier transform in the $\bar\lambda$ variable, which is an automorphism of representation (\ref{3oct7}). Without loss of generality we will focus on the first branch. 

Considering polynomial $\gamma_1$, we find the following spectrum of lowest-weight states 
\begin{equation}
\label{1sep11}
i \bar J_{\dot 1\dot 2}\delta(\bar\lambda^{\dot 1})\big(\bar\lambda^{\dot 2}\big)^k = \delta(\bar\lambda^{\dot 1})\left(k+1 \right)\big(\bar\lambda^{\dot 2}\big)^k,
\end{equation}
where $k\ge 0$. Accordingly, eigenvalues of $i \bar J_{\dot 1\dot 2}$ for the lowest-weight states are equal to $1,2,\dots$, which matches the spectrum we found previously for the tensor product of two singleton representations. 
We proceed further to the remaining states in the representation. By applying the raising operator we find
\begin{equation}
\label{1sep13}
(\bar J_{\dot 1\dot 1})^n \delta(\bar\lambda^{\dot 1}) \big(\bar\lambda^{\dot 2}\big)^k=
\left(-2i \bar\lambda^{\dot 2}\frac{\partial}{\partial \bar\lambda^{\dot 1}} \right)^n \delta(\bar\lambda^{\dot 1}) \big(\bar\lambda^{\dot 2}\big)^k
=(-2i)^n \delta^{(n)}(\bar\lambda^{\dot 1})\big(\bar\lambda^{\dot 2}\big)^{n+k}.
\end{equation}

It is now not hard to find a map that relates (\ref{3oct6}) and (\ref{1sep13}). Indeed, each factor of $a_1+ia_2$ in (\ref{3oct6}) goes over to a factor of $\bar\lambda^{\dot 2}$ in (\ref{1sep13}). Similarly, each factor of $a_1-ia_2$ in (\ref{3oct6}) generates a derivative of $\delta(\bar\lambda^{\dot 1}) $ in (\ref{1sep13}). The latter map can be implemented by first replacing $a_1-ia_2$ with some $\bar\mu$, which is then Fourier transformed to $\bar\lambda^{\dot 1}$. By putting these observations together, we find that the map between the two representations reads
\begin{equation}
\label{1sep15}
C_{\lambda_1,\lambda_2}(a_1,a_2)=\int d\bar\lambda^{\dot 1}e^{-i\frac{\alpha}{2}(a_1-ia_2)\bar\lambda^{\dot 1}}\Phi(\lambda;\bar\lambda^{\dot 1},\frac{1}{\alpha}(a_1+ia_2)).
\end{equation}
Here we added an extra undetermined variable $\alpha$, which accounts for the fact, that different irreducible representations, characterised by different lowest weights, can be intertwined with independent relative factors. 
Equation (\ref{1sep15}) implies that the intertwining kernel (\ref{3oct2}) so far, has been  fixed to be of the form
\begin{equation}
\label{1sep17}
k_\alpha(a_1,a_2,\bar\lambda^{\dot 1},\bar\lambda^{\dot 2}) = e^{-i\frac{\alpha}{2}(a_1-ia_2)\bar\lambda^{\dot 1}}
\delta(\bar\lambda^{\dot 2}-\frac{1}{\alpha}(a_1+ia_2)).
\end{equation}
The most general intertwining kernel, which is consistent with $\bar J$ symmetry, can be obtained from (\ref{1sep17}) by integration with an arbitrary weight $w(\alpha)$.
Having derived $k$ from the lowest weight considerations, we checked explicitly that it does satisfy (\ref{3oct3}) for $\bar J$ generators.

\subsection{Intertwining $P$ and $L$}

We now move on to the requirement of the invariance of $k$ under deformed translations. Explicitly one has 
\begin{equation}
\label{3oct8}
\begin{split}
P_{\alpha\dot 1} = \lambda_\alpha \bar\lambda^{\dot 2}, \qquad P_{\alpha\dot 2} = -\lambda_\alpha \bar\lambda^{\dot 1}
\end{split}
\end{equation}
in the higher-spin representation, while 
\begin{equation}
\label{3oct9}
\begin{split}
P_{\alpha\dot 1}=-\frac{1}{\sqrt{2}} \left( \lambda_{\alpha,1} a_1+  \lambda_{\alpha,2} a_2\right), \qquad  P_{\alpha\dot 2}=\frac{i}{\sqrt{2}}\left( \lambda_{\alpha,1 }\frac{\partial}{\partial a_1} +\lambda_{\alpha,2} \frac{\partial}{\partial a_2} \right)
\end{split}
\end{equation}
for the tensor product of singletons.
The invariance condition (\ref{3oct3}) for $P_{\alpha \dot 1}$ then leads to 
\begin{equation}
\label{1sep27}
\left[-\frac{1}{\sqrt{2}}(\lambda_{\alpha,1}a_1+\lambda_{\alpha,2} a_2)-\lambda_\alpha \bar\lambda^{\dot 2}\right]k_\alpha=0.
\end{equation}
With $k$ given in (\ref{1sep17}), this entails
\begin{equation}
\label{1sep28}
\lambda_{\alpha,1}=\frac{\sqrt{2}}{\alpha}\lambda_\alpha, \qquad 
\lambda_{\alpha,2}=i\frac{\sqrt{2}}{\alpha}\lambda_\alpha.
\end{equation}

As it was anticipated above, we find that $\lambda$'s of massless higher-spin fields and of two singletons, indeed, determine each other and, thus, the Flato-Fronsdal theorem in flat space works at fixed $\lambda$.
Imposing $P_{\alpha \dot 2}$ invariance is not necessary as it follows from $\bar J$ and $P_{\alpha \dot 1}$ invariance. Besides that, invariance with respect to $L$ also follows, as $L$ is generated from the commutator of two translations (\ref{19sep3}).

To summarise, we found that the intertwining map (\ref{3oct2}) between a pair of singleton representations and the massless higher-spin fields reads
\begin{equation}
\label{3oct2xx1}
C_{\frac{\sqrt{2}}{\alpha}\lambda,i\frac{\sqrt{2}}{\alpha}\lambda}(a_1,a_2)=v(\lambda)\int d\bar\lambda^{\dot 1}d\bar\lambda^{\dot 2}
e^{-i\frac{\alpha}{2}(a_1-ia_2)\bar\lambda^{\dot 1}}
\delta(\bar\lambda^{\dot 2}-\frac{1}{\alpha}(a_1+ia_2))
\Phi(\lambda;\bar\lambda^{\dot 1},\bar\lambda^{\dot 2}),
\end{equation}
where $v(\lambda)$ is an arbitrary function, which accounts for the fact that the Flato-Fronsdal theorem in flat space
works independently at different  $\lambda$. If, in addition, invariance with respect to $J$ (\ref{3oct2x01}) is imposed, then $v(\lambda)$ is constant. This is what we will assume in the following. 
Besides that, the kernel in (\ref{3oct2xx1}) features an undetermined parameter $\alpha$, which for given wave functions of singletons contributes to the higher-spin wave function as $\Phi(\alpha^{-1}\lambda,\alpha \bar\lambda)$.
Thus, the homogeneity degree in $\alpha$ counts the helicity of the higher-spin field and the ambiguity in $\alpha$ is related to the ambiguity to rescale fields of different helicities. This ambiguity is unphysical and was totally expected. Below we will keep $\alpha$ arbitrary. Still, it can be fixed from some additional considerations, such as the requirement that the intertwining kernel is higher-spin invariant, which we discuss in Appendix \ref{app:a}.

Finally, we remark that the map  defined by (\ref{3oct2xx1})  is, clearly, invertible, thus, the tensor product of two singletons is, indeed, equivalent to the tower of massless higher-spin representations. 

\subsection{Products of irreducible representations}
\label{sec3.3}

In this section we will give more refined versions of the flat space Flato-Fronsdal theorem, which take into account symmetry constraints
(\ref{23nov1}) making the singleton representation irreducible. 

We start by considering the tensor product of two scalar singletons. To this end, we start from the kernel of (\ref{3oct2xx1})
\begin{equation}
\label{23nov2}
K(\lambda,\bar\lambda |a_1,\lambda_1;a_2,\lambda_2)\equiv \delta^2 (\lambda_1 - \frac{\sqrt{2}}{\alpha}\lambda )\delta^2 (\lambda_2 - i\frac{\sqrt{2}}{\alpha}\lambda)e^{-i\frac{\alpha}{2}(a_1-ia_2)\bar\lambda^{\dot 1}}
\delta(\bar\lambda^{\dot 2}-\frac{1}{\alpha}(a_1+ia_2)),
\end{equation}
which we supplemented with additional delta functions making the $\lambda$-dependence explicit. For the tensor product of two scalar singletons this kernel should be projected in a way that it becomes even in (\ref{8oct1}) for each singleton leg. This leads to
\begin{equation}
\label{23nov3}
\begin{split}
&K^{{\rm scalar}\times{\rm scalar}}(\lambda,\bar\lambda |a_1,\lambda_1;a_2,\lambda_2)\\
& \qquad\qquad\qquad=\frac{1}{4}\Big(K(\lambda,\bar\lambda |a_1,\lambda_1;a_2,\lambda_2)+K(\lambda,\bar\lambda |-a_1,-\lambda_1;a_2,\lambda_2)\\
&\quad \qquad\qquad\qquad+K(\lambda,\bar\lambda |a_1,\lambda_1;-a_2,-\lambda_2)+
K(\lambda,\bar\lambda |-a_1,-\lambda_1;-a_2,-\lambda_2) \Big).
\end{split}
\end{equation}
It then straightforward to check that 
\begin{equation}
\label{23nov4}
K^{{\rm scalar}\times{\rm scalar}}(-\lambda,-\bar\lambda |a_1,\lambda_1;a_2,\lambda_2) =K^{{\rm scalar}\times{\rm scalar}}(\lambda,\bar\lambda |a_1,\lambda_1;a_2,\lambda_2).
\end{equation}
This implies that the tensor product of scalar singletons decomposes into the sum of bosonic higher-spin  fields
\begin{equation}
\label{23nov5}
\text{scalar}\times{\text{scalar}} = \text{bosonic HS}.
\end{equation}
In a similar manner we find that 
\begin{equation}
\label{23nov6}
\begin{split}
\text{spinor}\times{\text{spinor}} &= \text{bosonic HS},\\
\text{scalar}\times{\text{spinor}} &= \text{fermionic HS}.
\end{split}
\end{equation}

Another option is to consider symmetric tensor products. For the product of scalar singletons, the associated kernel is
\begin{equation}
\label{23nov7}
\begin{split}
&K^{{\rm Sym}({\rm scalar}\times{\rm scalar})}(\lambda,\bar\lambda |a_1,\lambda_1;a_2,\lambda_2)\\
&\qquad \qquad =\frac{1}{2}\left(K^{{\rm scalar}\times{\rm scalar}}(\lambda,\bar\lambda |a_1,\lambda_1;a_2,\lambda_2)+K^{{\rm scalar}\times{\rm scalar}}(\lambda,\bar\lambda |a_2,\lambda_2;a_1,\lambda_1)\right).
\end{split}
\end{equation}
It is straightforward to check that 
\begin{equation}
\label{23nov8}
K^{{\rm Sym}({\rm scalar}\times{\rm scalar})}(-i\lambda,i\bar\lambda |a_1,\lambda_1;a_2,\lambda_2) =K^{{\rm Sym}({\rm scalar}\times{\rm scalar})}(\lambda,\bar\lambda |a_1,\lambda_1;a_2,\lambda_2),
\end{equation}
which means that the symmetric product of two scalar singletons contains only even-spin fields 
\begin{equation}
\label{23nov9}
{\rm Sym}(\text{scalar}\times{\text{scalar}}) = \text{even bosonic HS}.
\end{equation}
Analogously, one finds that 
\begin{equation}
\label{23nov10}
{\rm Sym}(\text{spinor}\times{\text{spinor}}) = \text{odd bosonic HS}.
\end{equation}

Finally, we would like to note that the realisation of the flat space singleton as in (\ref{24aug2}) has some common features with the standard oscillatorial realisation of $so(3,2)$  singletons, while (\ref{3oct1}), (\ref{3oct2x01}) can be obtained by a smooth flat limit from the analogous realisaion of massless higher-spin fields in the AdS space.
Besides that, the flat space chiral higher-spin algebra is related to the higher-spin algebra in AdS by a smooth contraction.
 Similarly, the flat space kernel that we have found is reminiscent of the one that can be extracted from \cite{Vasiliev:2012vf}. Considering all these similarities and relations, it would be interesting to see whether our result can be obtained by a smooth  contraction of the kernel of \cite{Vasiliev:2012vf}.

\section{Higher-spin amplitudes from the dual theory}
\label{sec:4}

In the previous section we established a flat space version of the Flato-Fronsdal theorem. We also derived an explicit kernel that intertwines the higher-spin wave function with the wave function of the tensor product of two singleton fields. In the present section we will use this intertwining kernel to compute the simplest higher-spin amplitudes from the correlators of bilinear operators in the theory of flat space singletons. As in the simplest example of the higher-spin holography 
in the AdS space, we will assume that the boundary theory is free. This means that the correlators in this theory can be computed simply by performing the Wick contractions and, thus,  are expressed in terms of two-point functions of singletons.  

\subsection{Two-point function of singletons}

We start by deriving the two point function of singletons. As in the previous analysis, it will be fixed by requiring invariance with respect to the deformed Poincare algebra. 

Let $g_{\lambda_1,\lambda_2}(a_1,a_2)$ be the two-point function of the singleton. Each of the legs of the two-point function transforms in the representation, which is dual to  (\ref{24aug2}) or, in other words, operators in  (\ref{24aug2}) need to be integrated by parts.
Then invariance with respect to $\bar J$ requires
\begin{equation}
\label{1sep29}
\begin{split}
(a_1^2+a_2^2)g_{\lambda_1,\lambda_2}(a_1,a_2)=0, \\
-\left(\frac{\partial^2}{\partial a_1^2}+\frac{\partial^2}{\partial a_2^2} \right)g_{\lambda_1,\lambda_2}(a_1,a_2)=0,\\
i\left(a_1\frac{\partial}{\partial a_1}+a_2\frac{\partial}{\partial a_2}+1 \right)g_{\lambda_1,\lambda_2}(a_1,a_2)=0.
\end{split}
\end{equation}
The first of these equations implies 
\begin{equation}
\label{1sep30}
g_{\lambda_1,\lambda_2}(a_1,a_2)=k_1(a_1+ia_2) \delta(a_1-i a_2)+k_2(a_1-ia_2)\delta(a_1+ia_2),
\end{equation}
where $k_1$ and $k_2$ are arbitrary functions. Combining this with the remaining two equations, we find that 
\begin{equation}
\label{1sep31}
g_{\lambda_1,\lambda_2}(a_1,a_2)= \beta_1 \delta(a_1-i a_2)+\beta_2 \delta(a_1+ia_2).
\end{equation}

Next, we require that deformed translations leave the two-point function invariant. This leads to 
\begin{equation}
\label{1sep32}
-\frac{1}{\sqrt{2}}(\lambda_{\alpha,1} a_1+\lambda_{\alpha, 2}a_2)g_{\lambda_1,\lambda_2}(a_1,a_2)=0.
\end{equation}
Together with (\ref{1sep31}), this entails
\begin{equation}
\label{1sep34}
\begin{split}
g_{\lambda_1,\lambda_2}(a_1,a_2)&=g^1_{\lambda_1,\lambda_2}(a_1,a_2)+g^2_{\lambda_1,\lambda_2}(a_1,a_2),\\
g^1_{\lambda_1,\lambda_2}(a_1,a_2)&= \beta_1\delta^2(\lambda_{\alpha,1}+i\lambda_{\alpha,2})\delta(a_1+i a_2),\\
g^2_{\lambda_1,\lambda_2}(a_1,a_2)&=\beta_2 \delta^2(\lambda_{\alpha,1}-i\lambda_{\alpha,2})\delta(a_1-i a_2).
\end{split}
\end{equation}
The remaining generators leave the two-point function invariant automatically. 
One can consider two-point function of singletons with fixed parity (\ref{8oct1}), which will relate $\beta_1$ and $\beta_2$. For simplicity, below we consider $\beta_1=1$, $\beta_2=0$.

\subsection{Two-point amplitude of higher-spin fields}

We will now use the two-point functions of flat space singletons just derived to compute the two-point amplitude of massless higher-spin fields. 
To this end, as in the usual CFT relevant for the AdS space holography,  we consider the two-point correlator of single trace operators\footnote{Vector $O(N)$ indices can be added in a straightforward manner.}
\begin{equation}
\label{4oct1}
\langle : \varphi_1(a_1)\varphi_2(a_2): \; : \varphi_3(a_3)\varphi_4(a_4):\rangle,
\end{equation}
where $\varphi(a)$ is the operator of the singleton field localised at $a$.
Then, employing the intertwining kernels we derived previously,  $:\varphi_1\varphi_2:$ and $:\varphi_3 \varphi_4:$ will be converted to the higher-spin representation and (\ref{4oct1}) will be reinterpreted as the two-point amplitude of massless higher-spin fields in the Minkowski space. 

As we mentioned before, we assume that the dual theory is free, thus (\ref{4oct1}) is computed by performing the Wick contractions. There are two inequivalent ways to perform the Wick contractions in (\ref{4oct1})
\begin{equation}
\label{4oct2}
\begin{split}
&\langle : \varphi_1(a_1)\varphi_2(a_2):\; : \varphi_3(a_3)\varphi_4(a_4):\rangle \\ 
&\qquad \qquad \qquad \qquad =g_{\lambda_1,\lambda_3}(a_1,a_3) g_{\lambda_2,\lambda_4}(a_2,a_4)+
g_{\lambda_1,\lambda_4}(a_1,a_4)g_{\lambda_2,\lambda_3}(a_2,a_3).
\end{split}
\end{equation}

For definiteness, let us focus on the second term. We need to find a configuration of $\lambda$ for the singletons in such a way that $g_{\lambda_1,\lambda_4}(a_1,a_4)$, $g_{\lambda_2,\lambda_3}(a_2,a_3)$ as well as the intertwining kernels to higher-spin fields are all non-trivial. One such configuration is
\begin{equation}
\label{1sep35}
\lambda_1=\lambda, \quad \lambda_2=i \lambda, \quad \lambda_3=\lambda, \quad \lambda_4 = i\lambda,
\quad \lambda_a = \frac{\alpha}{\sqrt{2}}\lambda, \quad \lambda_b = \frac{\alpha}{\sqrt{2}}\lambda.
\end{equation}
Here $\lambda_a$ is the spinor for the higher-spin field arising as the tensor product of singletons $1$ and $2$, while $\lambda_b$ is the spinor for the higher-spin field associated with singletons $3$ and $4$. For values of $\lambda$ as indicated in (\ref{1sep35}) one has the first type of two-point functions (\ref{1sep34}) --
$g^1_{\lambda_1,\lambda_4}(a_1,a_4)$ and $g^1_{\lambda_2,\lambda_3}(a_2,a_3)$ -- non-vanishing.

The $\bar\lambda$-dependent part of the two-point amplitude of higher-spin fields then reads
\begin{equation}
\label{4oct3}
\begin{split}
A^{\rm r}(\bar\lambda_a,\bar\lambda_b)&= \int da_1 da_2 da_3 da_4  k_\alpha(a_1,a_2,\bar\lambda_a^{\dot 1},\bar\lambda_a^{\dot 2})\\
 & \qquad k_\alpha(a_3,a_4,\bar\lambda_b^{\dot 1},\bar\lambda_b^{\dot 2})g^1_{\lambda_1,\lambda_4}(a_1,a_4)g^1_{\lambda_2,\lambda_3}(a_2,a_3)\\
&=\int da_1 da_2 da_3 da_4 
e^{-i\frac{\alpha}{2}(a_1-ia_2)\bar\lambda_a^{\dot 1}}
\delta(\bar\lambda_a^{\dot 2}-\frac{1}{\alpha}(a_1+ia_2))\\
&\qquad e^{-i\frac{\alpha}{2}(a_3-ia_4)\bar\lambda_b^{\dot 1}}
\delta(\bar\lambda_b^{\dot 2}-\frac{1}{\alpha}(a_3+ia_4))\delta(a_1+ia_4)\delta(a_3+i a_2)\\
&= 2\pi \delta^2(\bar \lambda_a +\bar\lambda_b).
\end{split}
\end{equation}
Computation of integrals in (\ref{4oct3}) proceeds in a straightforward manner: three integrals are lifted by delta functions, while the remaining one is
the Fourier transform of an exponent.

Keeping in mind that $\lambda_a=\lambda_b$, see (\ref{1sep35}), we find that the complete amplitude is
\begin{equation}
\label{4oct4}
A(\lambda_a,\bar\lambda_a,\lambda_b,\bar\lambda_b)= 2\pi \delta^2( \lambda_a -\lambda_b) \delta^2(\bar \lambda_a +\bar\lambda_b).
\end{equation}
This result is consistent with \cite{Ponomarev:2022atv} in the sense that up to an irrelevant factor of $2\pi$ it agrees with the canonically normalised Wightman functions of massless higher-spin fields  summed over spins. 

\subsection{Three-point amplitude of higher-spin fields}

In a similar manner, we compute the three-point amplitude of higher-spin fields from the dual theory. To this end, we consider a correlator
\begin{equation}
\label{4oct5}
\begin{split}
&\langle : \varphi_1(a_1)\varphi_2(a_2): \; : \varphi_3(a_3)\varphi_4(a_4):  \; : \varphi_5(a_5)\varphi_6(a_6):\rangle\\
&\qquad \qquad \qquad \qquad  =g_{\lambda_2,\lambda_3}(a_2,a_3) g_{\lambda_4,\lambda_5}(a_4,a_5)g_{\lambda_6,\lambda_1}(a_6,a_1)+\dots.
\end{split}
\end{equation}
Focusing on the explicit term in (\ref{4oct5}), we choose a configuration of $\lambda$'s in a way that first branches of solution (\ref{1sep34}) for the two-point function are non-trivial. In addition, it is required that the singleton pairs $(1,2)$, $(3,4)$ and $(5,6)$ can be mapped to  higher-spin fields, the latter  labelled with $\lambda_a$, $\lambda_b$ and $\lambda_c$ respectively. In particular, we can choose
\begin{equation}
\label{1sep37}
\lambda_1=\lambda, \quad \lambda_2=i \lambda, \quad \lambda_3=\lambda, \quad \lambda_4 = i\lambda,
\quad \lambda_5 = \lambda, \quad \lambda_6 = i\lambda
\end{equation}
and 
\begin{equation}
\label{1sep38}
 \lambda_a = \frac{\alpha}{\sqrt{2}}\lambda, \quad \lambda_b =\frac{\alpha}{\sqrt{2}} \lambda, \quad \lambda_c=\frac{\alpha}{\sqrt{2}}\lambda.
\end{equation}

Then, the $\bar\lambda$-dependent part of the three-point amplitude of higher-spin fields reads
\begin{equation}
\label{4oct6}
\begin{split}
A^{\rm r}(\bar\lambda_a,\bar\lambda_b,\bar\lambda_c)&= \int da_1 da_2 da_3 da_4 da_5 da_6  k_\alpha(a_1,a_2,\bar\lambda_a^{\dot 1},\bar\lambda_a^{\dot 2})k_\alpha(a_3,a_4,\bar\lambda_b^{\dot 1},\bar\lambda_b^{\dot 2})\\
& 
\qquad k_\alpha(a_5,a_6,\bar\lambda_c^{\dot 1},\bar\lambda_c^{\dot 2})
g^1_{\lambda_2,\lambda_3}(a_2,a_3)g^1_{\lambda_4,\lambda_5}(a_4,a_5)g^1_{\lambda_6,\lambda_1}(a_6,a_1)\\
&=\int da_1 da_2 da_3 da_4  da_5 da_6
e^{-i\frac{\alpha}{2}(a_1-ia_2)\bar\lambda_a^{\dot 1}}
\delta(\bar\lambda_a^{\dot 2}-\frac{1}{\alpha}(a_1+ia_2))\\
&\qquad e^{-i\frac{\alpha}{2}(a_3-ia_4)\bar\lambda_b^{\dot 1}}
\delta(\bar\lambda_b^{\dot 2}-\frac{1}{\alpha}(a_3+ia_4))
e^{-i\frac{\alpha}{2}(a_5-ia_6)\bar\lambda_c^{\dot 1}}
\\
&\qquad \delta(\bar\lambda_c^{\dot 2}-\frac{1}{\alpha}(a_5+ia_6))\delta(a_3+i a_2)\delta(a_5+i a_4)\delta(a_1+i a_6)\\
&= 2\pi |\alpha| \delta^2(\bar \lambda_a +\bar\lambda_b+\bar\lambda_c)e^{\frac{i\alpha^2}{2}(-\bar\lambda^{\dot 2}_c \bar\lambda^{\dot 1}_b+\bar\lambda^{\dot 2}_b\bar\lambda^{\dot 1}_c)}.
\end{split}
\end{equation}
Again, the computation of integrals is very straightforward.

Reinstating the $\lambda$ dependence, which is defined by (\ref{1sep38}), we find
\begin{equation}
\label{4oct7}
A(\lambda_a,\bar\lambda_a,\lambda_b,\bar\lambda_b,
\lambda_c,\bar\lambda_c)= 2\pi  |\alpha| \delta^2( \lambda_a -\lambda_b)\delta^2( \lambda_b -\lambda_c) \delta^2(\bar \lambda_a +\bar\lambda_b+\bar\lambda_c)e^{\frac{i\alpha^2}{2}(-\bar\lambda^{\dot 2}_c \bar\lambda^{\dot 1}_b+\bar\lambda^{\dot 2}_b\bar\lambda^{\dot 1}_c)}.
\end{equation}
This result is consistent with the analysis of \cite{Ponomarev:2022atv} where the three-point amplitude in the chiral higher-spin theory  was summed over helicities on the external lines and brought to the form (\ref{4oct7}). It was also argued that up to a prefactor this result is fixed by the higher-spin symmetry of the theory.

\section{Conclusion}
\label{sec:5}

In the previous paper \cite{Ponomarev:2022ryp} we suggested a theory of flat space singletons and based on symmetries argued that it may provide a dual description of higher-spin gauge theories in the 4d Minkowski space. At the algebraic level this duality has many common features with the higher-spin AdS/CFT correspondence.  
At the same time, it departs substantially from the way  flat space holography is usually expected to work geometrically based on the analogy with the AdS space case -- namely, that the symmetry of the boundary theory should be the asymptotic symmetry of the Minkowski space, the BMS algebra. Instead, the symmetry of the dual theory in our approach is a certain deformation of the Poincare algebra, for which translations do not commute. For some representations -- such as for  massless higher-spin fields -- this deformation trivialises and, thus, one reproduce the usual Poincare algebra. At the same time, for the dual theory this deformation is necessarily non-trivial. 
This is what prevents us from regarding the dual theory as the boundary one and the duality, despite having some holographic flavour, is not, strictly speaking, holographic in the usual sense. 

In the present paper we provided substantial evidence that the duality suggested in \cite{Ponomarev:2022ryp} does work. We started by proving the flat space counterpart of the Flato-Fronsdal theorem \cite{Flato:1978qz}, that is that the tensor product of two singletons decomposes into the direct sum of massless higher-spin representations. This is an algebraic way of saying that the spectrum of  single-trace operators -- with the traces understood as in vector models -- in the theory of singletons matches the spectrum of fields in the Minkowski space. We would like to emphasise that, as argued in \cite{Ponomarev:2022ryp}, already this type of a statement about the spectrum is something that is impossible to achieve in  flat space holography, unless certain modifications -- such as the aforementioned deformation of the Poincare algebra -- are made. In other words, despite being just a basic statement at the level of the spectrum, the existence of the flat space Flato-Fronsdal theorem that we proved is a rather non-trivial fact, which, moreover, provides strong support for the approach to flat space holography we took. 

Besides that we computed the two- and three-point amplitudes of massless higher-spin fields in the 4d Minkowski space from the correlators in the dual theory by mimicking the standard steps from the AdS/CFT correspondence. The resulting amplitudes match the amplitudes in the chiral higher-spin theory in the form of \cite{Ponomarev:2022atv}. As argued there, higher-spin amplitudes are fixed by symmetries up to an independent overall factor for each number of external lines. Thus, this matching is to large extent controlled by higher-spin symmetries. Still, it allows one to fix the aforementioned overall factors, which were otherwise undetermined.

It should be remarked that in flat space only the chiral higher-spin theory  \cite{Metsaev:1991mt,Metsaev:1991nb,Ponomarev:2016lrm} is currently known how to define in a self-contained way, in particular, it has an explicit local action in the light-cone gauge. This theory is self-dual \cite{Ponomarev:2017nrr,Krasnov:2021nsq,Krasnov:2021cva}, which entails its integrability and the vanishing of its $n$-point amplitudes with $n >3$ \cite{Skvortsov:2018jea}. 
Despite in the present paper we did not compute holographic higher-point amplitudes  explicitly, it is clear that these are non-vanishing. This means that the higher-spin theory, which is dual to the theory of free flat space singletons is not the chiral higher-spin theory, but its certain extension. Such extensions do not have an independent bulk definition. The key reason for that is that numerous no-go theorems imply that massless higher-spin theories in flat space with non-trivial scattering do not exist unless locality is violated\footnote{This is a short and thus not quite accurate corollary of the no-go results, which are abundant and rather diverse. For precise statements and further details,  see e.g. reviews \cite{Bekaert:2010hw,Ponomarev:2022vjb}.}. In turn, by relaxing locality not only one risks to obtain theories displaying pathological physics, 
 one also looses a key consistency condition, which makes the typical perturbative approaches to the construction of higher-spin theories well-defined. Thus, the construction of higher-spin theories beyond the self-dual sector requires new guiding principles alternative to locality. From this perspective, we suggest the holographic approach of this paper as such an alternative. In other words, the holographic duality in the present case is, rather, a way to define a bulk theory than a conjecture that needs to be checked. In fact, to large extent, this also applies to the higher-spin holography in AdS, where an independent definition of the bulk theory faces difficulties related to locality, see e.g. discussions in \cite{Giombi:2009wh,Boulanger:2015ova,Bekaert:2015tva,Sleight:2017pcz,Ponomarev:2017qab,Didenko:2020bxd}. Such a holographic reconstruction may be used to define the flat space higher-spin theory $S$-matrix. Once the $S$-matrix is known one may, in principle, reconstruct the action of the theory in the light-cone gauge, by following the procedure detailed in \cite{Ponomarev:2016cwi}.

In the present paper, to describe the flat space singleton representation we used its  realisation in terms of a particular set of variables, which make the intertwining kernel of the flat space Flato-Fronsdal theorem particularly simple. At the same time, this description lacks some important features of the usual field theory, e. g. we cannot say much on how this representation can be extended off-shell, what is the natural manifold the field lives on and what is the associated action. More generally,  field theories with the underlying symmetry being the deformed Poincare algebra, that we encountered here,  have not been fully developed yet -- see e.g. \cite{Gomis:2017cmt} for recent discussions of models featuring the closely related Maxwell algebra as a symmetry. In fact, it is not naturally an algebra of vectors fields -- $L$ in (\ref{24aug2}) acts algebraically, which is more reminiscent of the Weyl transformations  -- thus, it is hard to interpret it as a space-time symmetry.
It would be interesting to explore this in future. 

The holographic construction that we presented here is essentially chiral. In particular, its global symmetry is a chiral algebra. 
To make contact with more usual theories it would be important to extend this approach to the parity-invariant setup. Besides that, the important role in our construction is played by $sl(2,\mathbb{C})$ spinors, which are particularly suited for dealing with massless fields in four dimensions. It would be interesting to see how this approach can be extended to massive fields and to other dimensions.

\acknowledgments

We would like to thank  V. Didenko and E. Skvortsov  for fruitful discussions on various subjects related to the paper. 
This research was partially completed at the workshop "Higher Spin Gravity and its Applications" supported by the Asia Pacific Center for Theoretical Physics.
I would also like to thank the participants of the  workshop for interesting comments.

\appendix

\section{Fixing $\alpha$ from higher-spin symmetry}
\label{app:a}

One way to get rid of the ambiguity related to independent rescalings of  fields with different helicities is to require that the kernel (\ref{3oct2xx1}) is, in addition, invariant with respect to higher-spin symmetries. Indeed, higher-spin generators with spin greater than two mix fields of different helicities and, thus, higher-spin invariance constrains relative normalisations between fields with different spins. 

Below we require the  invariance of the intertwining kernel with respect to the spin-4 generator $(\bar J_{\dot 1\dot 1})^3$. This generator acts on the higher-spin fields via 
\begin{equation}
\label{3oct10}
(\bar J_{\dot 1\dot 1})^3 \Phi\equiv (\bar J_{\dot 1\dot 1}\;\bar\circ\; \bar J_{\dot 1\dot 1} \;\bar\circ\;\bar J_{\dot 1\dot 1})\;\bar\circ\; \Phi
-\Phi  \;\bar\circ\;  (\bar J_{\dot 1\dot 1}\;\bar\circ\; \bar J_{\dot 1\dot 1} \;\bar\circ\;\bar J_{\dot 1\dot 1}),
\end{equation}
where 
\begin{equation}
\label{3oct11}
\bar J_{\dot 1\dot 1} = \frac{1}{2} \lambda_\alpha \lambda_\alpha
\end{equation}
and the $\bar\circ$ product is defined by \cite{Ponomarev:2022ryp}
\begin{equation}
\label{19sep1}
(f\;\bar\circ\; g)(\lambda,\bar\lambda) = f(\lambda,\bar\lambda) {\rm exp}\left[i \epsilon^{\dot\alpha\dot\beta}\frac{\overleftarrow\partial}{\partial \bar\lambda^{\dot\alpha}} \frac{\overrightarrow\partial}{\partial \bar\lambda^{\dot\beta}} \right] g(\lambda,\bar\lambda). 
\end{equation}
Evaluating (\ref{3oct10}) explicitly, we find 
\begin{equation}
\label{1sep19}
(\bar J_{\dot 1\dot 1})^3 \Phi =\left(
-\frac{3}{2}i (\bar \lambda^{\dot 2})^5 \frac{\partial }{\partial \bar\lambda^{\dot 1}}+5i
 (\bar \lambda^{\dot 2})^3 \frac{\partial^3 }{\partial (\bar\lambda^{\dot 1})^3}-
\frac{3}{2}i \bar \lambda^{\dot 2}   \frac{\partial^5 }{\partial (\bar\lambda^{\dot 1})^5}\right)\Phi.
\end{equation}
At the same time, for the tensor product of singletons one has
\begin{equation}
\label{1sep20}
\bar J_{\dot 1\dot 1}^3 C = (a_1^6+a_2^6)C.
\end{equation}
This leads us to the equation for the kernel 
\begin{equation}
\label{1sep21}
\left(-\frac{3}{2}i (\bar \lambda^{\dot 2})^5 \frac{\partial }{\partial \bar\lambda^{\dot 1}}+5i
 (\bar \lambda^{\dot 2})^3 \frac{\partial^3 }{\partial (\bar\lambda^{\dot 1})^3}-
\frac{3}{2}i \bar \lambda^{\dot 2}   \frac{\partial^5 }{\partial (\bar\lambda^{\dot 1})^5} +a_1^6+a_2^6\right)k=0,
\end{equation}
which can be only satisfied for 
\begin{equation}
\label{1sep22}
(\alpha^2-2)(\alpha^2+2)=0.
\end{equation}

The structure of the higher-spin algebra is such that the closure of the spin-2 generators and spin-4 generator $(\bar J_{\dot 1\dot 1})^3$ involves all higher-spin generators of even spins. This implies that for $\alpha$ satisfying (\ref{1sep22}), the intertwining kernel that we constructed is invariant with respect to  the  subalgebra of the higher-spin algebra containing only even spin generators.

To fix the remaining ambiguity, one may  consider spin-3 generator $(\bar J_{\dot 1\dot 1})^2$ in a similar way. It acts as 
\begin{equation}
\label{3oct12}
(\bar J_{\dot 1\dot 1})^2 \Phi \equiv
[ \bar J_{\dot 1\dot 1}\circ  \bar J_{\dot 1\dot 1},\Phi]_{\bar \circ}
=
\left(-2i (\bar \lambda^{\dot 2})^3 \frac{\partial }{\partial \bar\lambda^{\dot 1}}-2i
 (\bar \lambda^{\dot 2}) \frac{\partial^3 }{\partial (\bar\lambda^{\dot 1})^3}\right)\Phi
\end{equation}
on higher-spin fields. We found that for $(\bar J_{\dot 1\dot 1})^2$ acting on the tensor product of two singletons via $a_1^4+a_2^4$ -- which is how it naturally acts according to the definitions --  the invariance condition for the kernel has no solutions. In other words, with the action of the higher-spin algebra defined as above, the kernel cannot be made invariant with respect to higher-spin generators of odd spins.

It seems natural to expect that the Flato-Fronsdal kernel in flat space can be made invariant with respect to all higher-spin generators. To achieve this, we note that the higher-spin algebra has an automorphism
\begin{equation}
\label{23xnov1}
T_{s} \to (-1)^s T_{s},
\end{equation}
where the lower label refers to the spin of the generator. In other words, the automorphism (\ref{23xnov1}) changes the signs of odd spin generators.

The existence of the aforementioned automorphism implies  that when trying to achieve invariance of the intertwining kernel under spin-3 symmetry, we are free to change signs of $(\bar J_{\dot 1\dot 1})^2$ at our convenience. In particular, if the action of the higher-spin algebra on the second singleton is supplemented with (\ref{23xnov1}), then the action of $(\bar J_{\dot 1\dot 1})^2$ on the tensor product of singletons becomes $a_1^4-a_2^4$. One can then see that the invariance of the kernel  leads to 
\begin{equation}
\label{23xnov2}
\alpha^2=2.
\end{equation}
Alternatively, when the action of the higher-spin algebra on the first singleton is corrected with (\ref{23xnov1}), then one finds 
\begin{equation}
\label{23xnov3}
\alpha^2=-2.
\end{equation}

For definiteness, we will choose the first option (\ref{23xnov2}), which leads to two possible values of $\alpha$ differing by the overall sign. As it is not hard to see, the effect of the change of the sign of $\alpha$ is that fermionic higher-spin fields change the sign, while bosonic higher-spin fields remain intact. Irreducible kernels, which we discussed in section \ref{sec3.3}, do not involve higher-spin bosons and fermions simultaneously. Thus, the effect of the sign change for $\alpha$ on irreducible kernels reduces to the change of their overall normalisation and, therefore, is immaterial. 

In summary, we have shown that by requiring invariance with respect to the higher-spin symmetry, the irreducible intertwining kernels of the flat space Flato-Fronsdal theorem are fixed up to an overall factor. It, still, seems somewhat unnatural that to achieve that the automorphism (\ref{23xnov1}) was necessary. It would be interesting to clarify this in future.

\bibliography{flatff}
\bibliographystyle{JHEP}

\end{document}